\documentclass[review]{elsarticle}
\usepackage{amsmath,amssymb}
\usepackage{lineno,hyperref}
\usepackage{graphicx}
\modulolinenumbers[5]
\journal{Journal of \LaTeX\ Templates}

\begin{document}
\begin{frontmatter}
\title{The role of solid surface in bubble formation and detachment at a submerged orifice}
\author[mymainaddress]{Wenbiao Jiang\corref{mycorrespondingauthor}\fnref{myfootnote}}
\fntext[myfootnote]{Permanent Email address: jiangwb43@sina.com; ORCID: orcid.org/0000-0003-3978-989X}
\cortext[mycorrespondingauthor]{Corresponding author}
\ead{wenbiao.jiang@centralesupelec.fr}
\address[mymainaddress]{Laboratoire de G\'{e}nie des Proc\'{e}d\'{e}s et Mat\'{e}riaux, CentraleSup\'{e}lec, Universit\'{e} Paris-Saclay, 3 Rue Joliot Curie, 91190 Gif-sur-Yvette, France}
\begin{abstract}
	\begin{center}
		\includegraphics{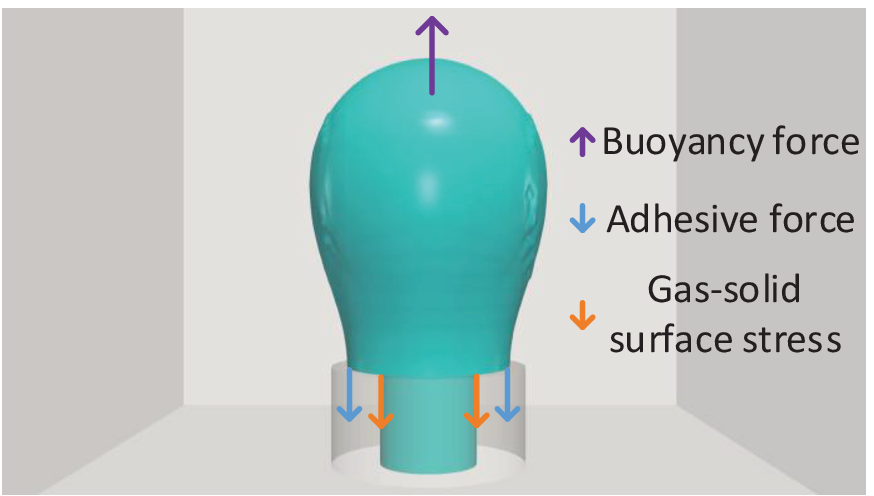}
	\end{center}
	
	Bubble formation and detachment at a submerged orifice exists widely in both daily life and academic research, and the influence of solid surface on bubble formation and detachment has also been extensively investigated. For example, it has been observed that the bubble on a hydrophobic surface is larger than the one on a hydrophilic surface. However, this phenomenon is not sufficiently explained in existing literature. To explain it, we redefine the capillary force at the orifice edge and the adhesive force between bubble and solid. Subsequently, we are able to establish an approach that quantitatively explains the relation between bubble volume and surface wettability. In addition, we also suggest an alternative mechanism of bubble pinch-off. In our opinion, the formation of bubble neck during pinch-off may be due to three-phase equilibrium at the orifice edge, and the breakup of bubble neck may result from the stretching action of the upward buoyancy force and the downward capillary force.
\end{abstract}
		
\end{frontmatter}

\section{Introduction}
Children enjoy making bubbles in beverages with a straw, an air pump blows bubbles into an aquarium for the fish to breathe better, bubbles are injected into an airlift reactor to react with liquid, all these familiar phenomena can be regarded as bubble formation and detachment at a submerged orifice. At low gas flow rate, bubble formation is determined mainly by the upward buoyancy force $F_B$ and the downward capillary force $F_C$ \cite{benzing1955low}, and the bubble can grow as long as $F_B$ is smaller than $F_C$. Once $F_B$ attains equilibrium with $F_C$, the bubble will detach due to the unbalanced forces with further bubble growth. However, the definition of capillary force at the orifice edge seems to be debatable, since the capillary force at the orifice edge was equated with the one inside a tube \cite{di2013experimental} \cite{lesage2013modelling} \cite{esfidani2017modeling}. Besides, a bubble possibly spreads over the horizontal solid surface \cite{gnyloskurenko2003influence}, and accordingly bubble volume changes with surface wettability. However, the existing theory can not give exact prediction about the variation of bubble volume with surface wettability.

By proposing alternative formulas of both capillary force and adhesive force, we explain the influence of wettablity on bubble volume. Meanwhile, we find that bubble formation and detachment are mainly related to the gas-solid and liquid-solid interfaces, to which previous research seems not to be given sufficient importance. Once we realize the genuine factors that determine bubble formation and detachment, we will be able to control the bubble volume and form, which is an important issue in the petrochemical industry, cosmetic industry, nuclear reactors, mineral processing, etc. \cite{rana2017towards}. In addition, we provide an alternative explanation for bubble pinch-off, which may be helpful to related research \cite{keim2006breakup}.

In this article, we establish another approach to explain bubble formation and detachment at low gas flow rate. To achieve this, we exclude the gas-liquid surface stress from the forces acting on the bubble as a whole. The gas-liquid surface stress is an internal force for the bubble as a whole, it balances with the pressure drop in the normal direction, yet it can not resist tangentially an external force, namely the buoyancy force. Instead, the forces to resist the buoyancy force are mainly the capillary force and the adhesive force. The capillary force is proportional to the orifice radius, and the adhesive force depends on wettability of the horizontal surface. Finally, we estimate the final bubble volume by equating the resisting forces and the buoyancy force. In this way, we can reason how the bubble volume varies with the orifice radius and wettability of the horizontal solid surface.
\section{Theory}
\begin{figure}[t]
	\centering
	\includegraphics[width=0.8\textwidth]{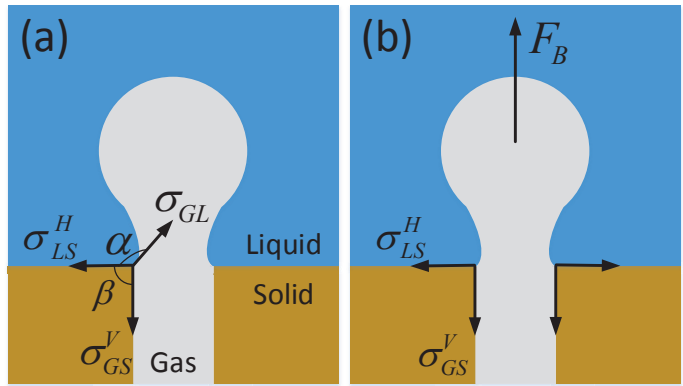}
	\caption{A bubble that does not spread beyond the orifice edge. (a) The three surface stresses $\sigma$ act on the triple line located at the orifice edge. $\alpha$, $\beta$ are the angles between the interfaces, and $\beta$ is set to be $\pi/2$. (b) The gas-liquid surface stress $\sigma_{GL}$ does not appear in the forces acting on the bubble as a whole. The liquid-solid surface stress $\sigma_{LS}^H$ pulls the bubble outwards and may make the bubble spread over the horizontal surface. The  gas-solid surface stress $\sigma_{GS}^V$ resists the buoyancy force $F_B$.}
	\label{forces}
\end{figure}

To simplify the physical model, we only investigate the situations with a low gas flow rate. Therefore, the bubble growth can be regarded as a quasi-static process. Subsequently, we consider the situation where a bubble does not spread over the horizontal solid surface, as shown in Fig. \ref{forces}. In this case, a three-phase contact line is located at the orifice edge. Thus, referring to Fig. \ref{forces}(a), the surface stresses in equilibrium satisfy \cite{rowlinson2002molecular}
\begin{eqnarray}
\sigma_{LS}^H=-\sigma_{GL}\cos\alpha \label{lg} \\
\sigma_{GS}^V=\sigma_{GL}\sin\alpha \label{gs},
\end{eqnarray}
where $\alpha$ is the contact angle, $H$, $V$ stand for ``Horizontal'', ``Vertical'', respectively, and $G$, $L$, $S$ stand for ``Gas'', ``Liquid'', ``Solid'', respectively. In this way, $\sigma_{GL}$ is the gas-liquid surface stress, $\sigma_{LS}^H$ is the surface stress at the horizontal liquid-solid interface and $\sigma_{GS}^V$ is the surface stress at the vertical gas-solid interface.

In addition, both of the horizontal and vertical solid surfaces satisfy
\begin{eqnarray}
\sigma_{GL}\cos\theta_H=\sigma_{GS}^H-\sigma_{LS}^H \label{hss}\\
\sigma_{GL}\cos\theta_V=\sigma_{GS}^V-\sigma_{LS}^V, \label{vss}
\end{eqnarray}
known as the Young equation, where $\theta$ is the contact angle on a perfectly flat and rigid surface. This angle also characterizes wettability of the surface: a small $\theta$ implies high wettabilty (hydrophilic surface), whereas a large $\theta$ signifies low wettability (hydrophobic surface). If the material of the horizontal and vertical surfaces is identical, from equations (\ref{lg}) (\ref{gs}) (\ref{hss}) (\ref{vss}), we have
\begin{equation}
\cos\left(\alpha-\frac{\pi}{4}\right)=\frac{\sqrt{2}}{2}\cos\theta,
\end{equation}
where $\theta=\theta_H=\theta_V$. As a result, we obtain $\pi/2<\alpha<\pi$ for $0<\theta<\pi$, this obtuse contact angle $\alpha$ is observed during bubble pinch-off \cite{burton2005scaling} \cite{thoroddsen2007experiments}. In other words, the initial stage of bubble pinch-off, namely the formation of the bubble neck, may result from the fact that the three phases seek to attain equilibrium at the orifice edge.

Fig. \ref{forces}(b) shows the forces acting on the bubble as a whole. Inside a tube, the capillary force is \cite{de2004capillarity}
\begin{equation}
F_C=\sigma_{GS}-\sigma_{LS}.
\end{equation}
However, liquid-solid surface stress is a horizontal force at the orifice edge, naturally, it does not appear in the vertical capillary force. Therefore, the capillary force at orifice edge is the gas-solid surface stress
\begin{equation}
F_C=\sigma_{GS}^V,
\end{equation}
which resists the buoyancy force $F_B$. During bubble detachment, upward $F_B$ and downward $\sigma_{GS}^V$ will stretch the bubble to break up, which may explain the observations from \cite{keim2006breakup}: ``In the last stages (of bubble pinch-off), the air appears to tear instead of pinch''. In horizontal direction, $\sigma_{LS}^H$ tends to pull the bubble onto the horizontal solid surface. Consequently, the bubble may spread over the horizontal surface. In this case, the buoyancy force is met with resistance mainly from two forces: the surface stress $\sigma_{GS}^V$ provided by the vertical gas-solid interface, and the adhesive force $F_a$ provided by the horizontal gas-solid interface. 
\begin{figure}[t]
	\centering
	\includegraphics[width=0.8\textwidth]{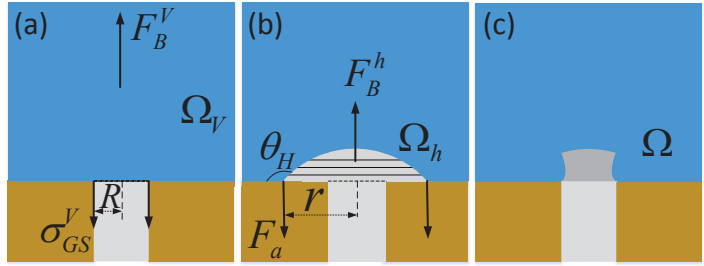}
	\caption{The method for calculating bubble volume. (a) $\Omega_V$ is the maximum gas volume that can be supported by the vertical gas-solid interface. (b) During bubble spreading, $\Omega_h$ is the gas volume that is held by the horizontal solid surface, its maximum value is $\Omega_H=(\Omega_h)_{max}$. (c) The bubble volume is considered as a sum of the previous two maximum volumes. Noticing that the overlapping volume represented by the dark region appears twice in the total volume, thus the theoretical bubble volume is a little larger than the actual one.}
	\label{volume}
\end{figure}

To calculate the maximum bubble volume $\Omega$, we artificially divide the bubble into two parts. A ``vertical part'' that is supported by the vertical gas-solid interface, and a ``horizontal part'' that adheres to the horizontal solid surface, as shown in Fig. \ref{volume}(a) and (b), respectively. Thus
\begin{equation}
\Omega=\Omega_V+(\Omega_h)_{max},
\label{vt}
\end{equation} 
where $\Omega_V$ is the maximum volume of the vertical part, $\Omega_h$ is the volume of the horizontal part, its maximum value is $\Omega_H=(\Omega_h)_{max}$.

By equating the corresponding buoyancy force $F_B^V=\Omega_V \Delta\rho g$ and the total surface stress $F_{GS}^V=2\pi R\sigma_{GS}^V$, we estimate $\Omega_V$ in Fig. \ref{volume}(a) as
\begin{equation}
\Omega_V=\frac{2\pi\sigma_{GS}^V}{\Delta\rho g}R,
\label{vv}
\end{equation}
where $R$ is the orifice radius, $\Delta\rho$ is the density difference between liquid and gas, $g$ is the gravitational acceleration.

To simplify the algebraic calculations, we suppose that the horizontal part has the shape of a spherical cap, thus
\begin{equation}
\Omega_h(\theta_H,r)=\frac{\pi(2+3\cos\theta_H-\cos^3\theta_H)r^3}{3\sin^3\theta_H},
\label{vdh}
\end{equation}
where $r$ is the spreading radius on the horizontal surface, as shown in Fig. \ref{volume}(b). The corresponding buoyancy force is $F_B^h=\Delta\rho g\Omega_h$. It is worth mentioning that the actual horizontal part is the spherical cap subtracted by the overlapping part, which is represented by the dark region in Fig. \ref{volume}(c). In other words, the actual $\Omega_h$ depends on the orifice radius $R$. Nevertheless, the spherical cap can approximately represent the horizontal part by assuming that $R$ is small enough, namely the overlapping volume is negligible.

To pull the horizontal part away from the horizontal surface, the required work per unit area is
\begin{equation}
w_H=\sigma_{GL}+\sigma_{LS}^H-\sigma_{GS}^H,
\label{wh}
\end{equation}
known as the work of adhesion \cite{israelachvili2011intermolecular}. Substituting equation (\ref{hss}) into (\ref{wh}), we have
\begin{equation}
w_H=\sigma_{GL}(1-\cos\theta_H).
\end{equation}
It has been found that the adhesive force per unit length is proportional to the work of adhesion \cite{samuel2011study} \cite{xue2014solid}, hence we obtain
\begin{equation}
F_a=2\pi rk_a\sigma_{GL}(1-\cos\theta_H),
\label{fa}
\end{equation}
where $k_a$ is a dimensionless coefficient relating the adhesive force to the work of adhesion. 
\begin{figure}[t]
	\centering
	\includegraphics[width=0.8\textwidth]{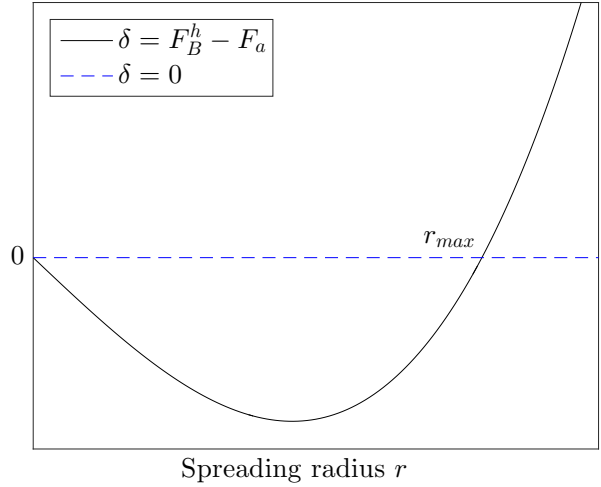}
	\caption{The difference between the buoyancy force $F_B^h$ and the adhesive force $F_a$ versus spreading radius $r$ for the horizontal part. $\delta=0$ indicates $F_B^h=F_a$, $\delta<0$ indicates $F_B^h<F_a$. And $r_{max}$ is the maximum spreading radius.}
	\label{diff}
\end{figure}

To evaluate the competition between the corresponding buoyancy force and the adhesive force, we calculate the difference
\begin{equation}
\delta=F_B^h-F_a=C\left[r^3-\frac{6k_a\sigma_{GL}}{\Delta\rho g}\frac{(1-\cos\theta_H)\sin^3\theta_H}{2+3\cos\theta_H-\cos^3\theta_H}r\right],
\end{equation}
where
\begin{equation}
C=\frac{\pi(2+3\cos\theta_H-\cos^3\theta_H)\Delta\rho g}{3\sin^3\theta_H}
\end{equation}
is positive for $0<\theta_H<\pi$. We plot $\delta$ versus $r$ in Fig. \ref{diff}, and we observe that for $r<r_{max}$ the buoyancy force is smaller than the adhesive force. The bubble continues to spread over the horizontal surface until the spreading radius attains its maximum value
\begin{equation}
r_{max}=\sqrt{\frac{6k_a\sigma_{GL}}{\Delta\rho g}\frac{(1-\cos\theta_H)\sin^3\theta_H}{2+3\cos\theta_H-\cos^3\theta_H}},
\label{rmax}
\end{equation} 
at which the buoyancy force $F_B^h$ equals the adhesive force $F_a$. For $r>r_{max}$, the buoyancy force is larger than the adhesive force, thus the bubble starts detaching.
\begin{figure}[t]
	\centering
	\includegraphics[width=0.8\textwidth]{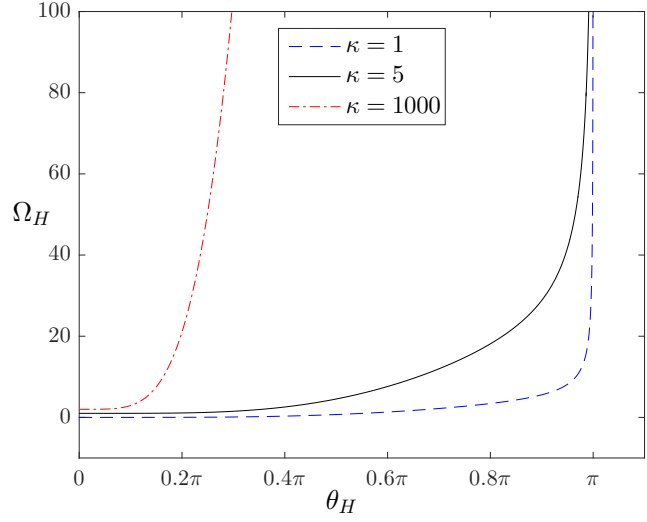}
	\caption{Function $\Omega_H=\kappa\left[\frac{(1-\cos\theta_H)^3\sin^3\theta_H}{2+3\cos\theta_H-\cos^3\theta_H}\right]^{\frac{1}{2}}$ for different $\kappa$. The curves are slightly staggered for better visibility of the flat tails.}
	\label{fk}
\end{figure}

The maximum volume of the horizontal part is estimated as $\Omega_H=\Omega_h(r_{max})$ for a fixed $\theta_H$. Together with equations (\ref{vdh}) (\ref{rmax}), we obtain
\begin{equation}
\Omega_H=\frac{\pi}{3}\left(\frac{6k_a\sigma_{GL}}{\Delta\rho g}\right)^{\frac{3}{2}}\left[\frac{(1-\cos\theta_H)^3\sin^3\theta_H}{2+3\cos\theta_H-\cos^3\theta_H}\right]^{\frac{1}{2}},
\label{vh}
\end{equation}
a function of wettability of the horizontal surface $\theta_H$. To analyze equation (\ref{vh}), we regard
\begin{equation}
\kappa=\frac{\pi}{3}\left(\frac{6k_a\sigma_{GL}}{\Delta\rho g}\right)^{\frac{3}{2}}.
\end{equation}
Subsequently, we plot $\Omega_H$ for different $\kappa$ in Fig. \ref{fk}, and we observe that $\Omega_H$ has a flat tail where $\Omega_H$ approximately equals zero. Furthermore, the size of this tail is determined by the parameter $\kappa$, a larger $\kappa$ implies a smaller tail. In the context of bubble formation, this flat tail indicates that a bubble can hardly spread over a relatively hydrophilic surface (with small $\theta_H$). Lin \textit{et al.} \cite{lin1994role} have described this phenomenon as: ``the bubble growth on the hydrophobic surface is no longer taking place at the edge of the orifice as the case of hydrophilic surfaces. In fact, the contact base of the bubble started to spread beyond the orifice edge as the orifice surface became more hydrophobic.''

Substituting equations (\ref{vv}) and (\ref{vh}) into (\ref{vt}), we obtain the total volume
\begin{equation}
\Omega(R,\theta_H)=\frac{2\pi\sigma_{GS}^V}{\Delta\rho g}R+\frac{\pi}{3}\left(\frac{6k_a\sigma_{GL}}{\Delta\rho g}\right)^{\frac{3}{2}}\left[\frac{(1-\cos\theta_H)^3\sin^3\theta_H}{2+3\cos\theta_H-\cos^3\theta_H}\right]^{\frac{1}{2}}.
\label{ttv}
\end{equation}
\begin{figure}[t]
	\centering
	\includegraphics[width=0.8\textwidth]{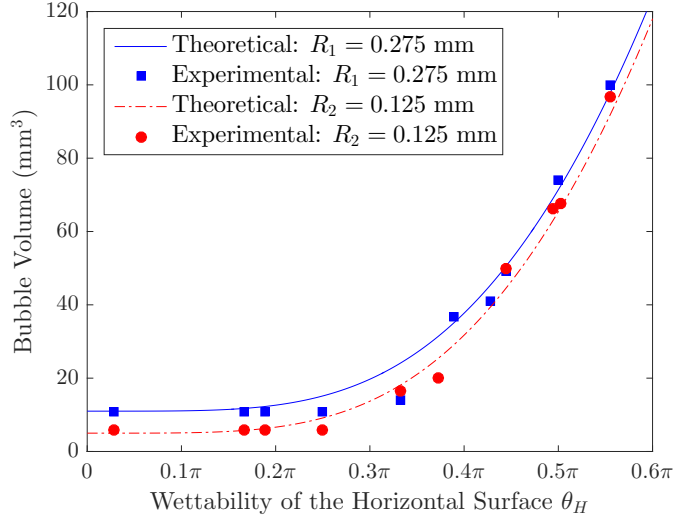}
	\caption{Bubble volume versus wettability: a comparison between the theoretical prediction (\ref{ttv}) and the experimental data from \cite{lin1994role}. The theory aligns well with the experiments for two orifice sizes, $R_1=0.275$ mm and $R_2=0.125$ mm.}
	\label{wett}
\end{figure}
Equation (\ref{ttv}) explains the following observations made by Lin \textit{et al.} \cite{lin1994role}: ``For orifice contact angles ($\theta_H$) between 0 and 55$^\circ$, the bubble volume is determined by the orifice diameter and independent of the magnitude of the contact angle. However, when the orifice contact angle exceeds the threshold value of 55$^\circ$, the air bubble volume increased with increasing contact angle due to spreading of the contact base.''

Inspecting equation (\ref{ttv}) we see that for small $\theta_H$, the total volume is determined by the orifice size in the first term of equation (\ref{ttv}), since the second term almost vanishes. Once the second term leaves the flat tail, it increases dramatically with increasing $\theta_H$. As a result, the contribution of the second term to the total volume will become much larger than the first term, naturally, as described by \cite{lin1994role}: ``the contribution of orifice size to the bubble volume is important only for hydrophilic surfaces.''.

Similar experimental results, i.e. constant bubble volume on hydrophilic surface and dramatical increase of bubble volume with contact angle on hydrophobic surface, are also been reported in other articles \cite{byakova2003influence} \cite{kukizaki2008effect} \cite{gnyloskurenko2003wettability} \cite{corchero2006effect} \cite{wesley2016influence}, they can also be explained in principle by equation (\ref{ttv}).

Fig. \ref{wett} shows a quantitative comparison between the theoretical prediction (\ref{ttv}) and the experimental data from \cite{lin1994role}. The constants in equation (\ref{ttv}) are determined as follows. We assume that $\Omega_H$ is 0 for $\theta_H=0.087$, consequently, from $\Omega(R_1=0.275\ mm,\theta_H=0.087)=11\ \text{mm}^3$, we have
\begin{equation}
\frac{2\pi \sigma_{GS}^V}{\Delta\rho g}\approx 40\ \text{mm}^2.
\label{c1}
\end{equation}
This is a constant, since the material of the vertical solid surface does not change in the experiments. Subsequently, by minimizing the mean square error between the theoretical and experimental values for $R_1=0.275$ mm, we obtain
\begin{equation}
\frac{\pi}{3}\left(\frac{6k_a\sigma_{GL}}{\Delta\rho g}\right)^{\frac{3}{2}}\approx 85.4\ \text{mm}^3,
\label{c2}
\end{equation}
which is also invariant by assuming that $k_a$ in equation (\ref{fa}) is a constant. With the constants obtained from $R_1=0.275$ mm, the established equation also predicts the experimental data correctly for the second orifice size, namely $R_2=0.125$ mm.

In addition, from equation (\ref{c1}), we have
\begin{equation}
\sigma_{GS}^V\approx 60\text{ mJ/m}^2,
\label{gsse}
\end{equation}
the surface energy of 304 stainless steel, i.e. the orifice material used in the experiments \cite{lin1994role}. The surface energy of 304 stainless steel in other literature is about 50$\sim$60 mJ/m$^2$ \cite{mantel1994influence} \cite{thongyai2005study}, which confirms the value (\ref{gsse}) deduced from our approach. This also validates our approach from another point of view.
\section{Acknowledgments}
W. Jiang gratefully acknowledges financial support from the China Scholarship Council (CSC 201504490092). Also, W. Jiang established the approach thanks to the helpful discussions with Prof. Fran\c cois Puel, Mr. Stefan Kinauer, Mr. Moustapha Diallo and Mr. Kaili Xie.
\section*{References}
\bibliographystyle{elsarticle-num}
\bibliography{Jiang_BubbleFormation}
\end{document}